\documentclass[12pt]{amsart}
\usepackage{graphicx}
\usepackage{epstopdf}
\RequirePackage{color}
\RequirePackage[colorlinks, urlcolor=my-blue,linkcolor=my-red,citecolor=my-green]{hyperref}
\definecolor{my-blue}{rgb}{0.0,0.0,0.6}
\definecolor{my-red}{rgb}{0.5,0.0,0.0}
\definecolor{my-green}{rgb}{0.0,0.5,0.0}
\definecolor{nicos-red}{rgb}{0.75,0.0,0.0}
\definecolor{light-gray}{gray}{0.6}
\definecolor{really-light-gray}{gray}{0.8}
\definecolor{sussexg}{rgb}{0.0,0.5,0.5}
\definecolor{sussexp}{rgb}{0.5,0.0,0.5}

\usepackage{geometry}                
\usepackage[parfill]{parskip}    

\usepackage{amsmath, amsthm, amssymb}
\usepackage{multirow,url}
\usepackage{verbatim}
\usepackage{amsfonts}
\usepackage{hyperref}
\usepackage{latexsym}
\usepackage{array}
\usepackage{subfig}

\newtheorem{theorem}{Theorem}[section]

\makeatletter
\newcommand{\addresseshere}{%
  \enddoc@text\let\enddoc@text\relax
}
\makeatother

\begin{document}

\title{Simulations on the combinatorial structure of D-optimal designs}

\author{Roberto Fontana}

\address{Department DISMA, Politecnico di Torino, Corso Duca degli Abruzzi 24, 10127 TORINO, Italy}
\email{roberto.fontana@polito.it}

\author{Fabio Rapallo}

\address{DISIT, Universit\`a del Piemonte Orientale, Alessandria, Italy}
\email{fabio.rapallo@uniupo.it}





\begin{abstract}
In this work we present the results of several simulations on main-effect factorial designs. The goal of such simulations is to investigate the connections between the $D$-optimality of a design and its geometrical structure. By means of a combinatorial object, namely the circuit basis of the design matrix, we show that it is possible to define a simple index that exhibits strong connections with the $D$-optimality.
\end{abstract}



\maketitle

\section{Introduction}
Many experimental situations call for standard designs, such as fractional factorials. However in many situations standard designs are not available, for example when not all combinations of the factor levels are feasible or resource limitations restrict the number of experiments that can be performed. In these non-standard situations $D$-optimal designs are often used \cite{mitchell1974computer}.

In the recent work \cite{jspi2014}, saturated fractions, which are designs where the number of points is equal to the number of estimable parameters of the model, have been characterized through the circuits of the model matrix. The key point of such theory is the identification of a fraction with a $\{0,1\}$-table where the points belonging to the fraction are denoted with 1 and the points outside the fraction are denoted with 0. Circuits are algebraic objects derived from the model matrix. We will recall the definition and the basic properties of circuits in the next section. The structure of saturated $D$-optimal designs in connection with the circuits has been studied in \cite{rimini2014d}.

Since the circuits yield major information on the $D$-optimality of saturated fractions, in this work we perform a simulation study for investigating the geometric structure of non-saturated $D$-optimal designs in connection with their circuits. We limit our analysis to main-effect models and we present some test cases dealing with both symmetric and asymmetric designs. From these examples, we argue that there are strong connections between the $D$-optimality and the circuits. We use {\tt Proc Optex} of {\tt SAS/QC} \cite{man3} for generating $D$-optimal designs and {\tt 4ti2} \cite{4ti2} for generating circuits.

{\tt Proc Optex} searches for optimal experimental designs in the following way. The user specifies an efficiency criterion, a set of candidate design points, a model and the size of the design to be found, and the procedure generates a subset of the candidate set so that the terms in the model can be estimated as efficiently as possible. There are several algorithms for searching for $D$-optimal designs. They have a common structure. They start from an initial design, randomly generated or user specified, and move, in a finite number of steps, to a better design. All of the search algorithms are based on adding points to the growing design and deleting points from a design that is too big. Main references to optimal designs include \cite{atkinson2007optimum}, \cite{goos2011optimal}, \cite{pukelsheim2006optimal}, \cite{rasch2011optimal}, \cite{shah1989theory} and \cite{wynn1970sequential}.

{\tt 4ti2} is a symbolic software which computes the circuits of a given integer-valued matrix. The use of software for Combinatorics and Computer Algebra inside statistical simulations usually leads to limitations in the size of the problems. The algorithms become actually unfeasible when the number of the design points grows, and our problem does not make exception. Therefore, we will restrict to small-sized examples.

It is worth noting that despite these computational limitations, we are able to consider a variety of examples, including both binary and multilevel designs, with an example of mixed-level design.

The paper is organized as follows. In Sect.~\ref{sec:1} we briefly describe the results of \cite{jspi2014} and in particular how saturated designs can be characterized in terms of the circuits of the relevant model matrix. In Sect.~\ref{sec:2} we recall some major results on the $D$-optimality of saturated fractions and we introduce our simulation study for main-effect models. In Sect.~\ref{sec:3} we present and discuss the results of our simulations, while in Sect.~\ref{sec:conclusion} we give some concluding remarks and some pointers to future work.

\section{Circuits, saturated designs, and $D$-optimality}
\label{sec:1}

In this section we recall the definition of circuits and we review their applications to Design of Experiments. A full account on circuits including some applications to Statistics is available in \cite{ohsugi:12}.

Given a model matrix $X$ on a full factorial design ${\mathcal D}$ with $K$ design points, an integer vector $f$ is in the kernel of $X^t$ if and only if $X^t f=0$. We denote by $A$ the transpose of $X$. Moreover, we denote by ${\mathrm{supp}}(f)$ the support of the integer vector $f$, i.e., the set of indices $j$ such that $f_j \ne 0$. Finally, the indicator vector of $f$ is the binary vector $(f_j \ne 0)$, where $( \cdot )$ is the indicator function. An integer vector $f$ is a circuit of $A$ if and only if:
\begin{itemize}
\item[(a)] $f \in \ker(A)$;

\item[(b)] there is no other integer vector $g \in {\ker(A)}$ such that $\mathrm{supp}(g) \subset \mathrm{supp}(f)$ and $\mathrm{supp}(g) \ne \mathrm{supp}(f)$.
\end{itemize}

The set of all circuits of $A$ is denoted by ${\mathcal C}_A=\{f_1, \ldots , f_L\}$, and is named as the circuit basis of $A$. It is known that ${\mathcal C}_A$ is always finite. The set ${\mathcal C}_A$ can be computed
through specific software. In our examples, we have used {\tt 4ti2} \cite{4ti2}. Notice that ${\mathcal C}_A$ is a special basis of $\ker(A)$ as vector space, and therefore the circuit basis is computed from the model matrix on the full factorial design ${\mathcal D}$. Thus, the circuit basis ${\mathcal C}_A$ depends only on the model, but not on the fraction. This remark is particularly useful when we use this theory in the definition of algorithms for finding optimal designs, since the computation of the circuit basis.

The connection between saturated fractions and circuits is given in the following theorem, to be found in \cite{jspi2014}. Remember that saturated fractions are fractions with the minimal number of points $p={\mathrm{rank}}(A)$ such that  all the $p$ independent parameters are estimable.

\begin{theorem} \label{mainthm}
A fraction ${\mathcal F} \subset {\mathcal D}$ with $p$ design points is a saturated fraction if and only if it does not contain any of the supports $\{\mathrm{supp}(f_1), \ldots,
\mathrm{supp}(f_L)\}$ of the circuits of $A=X^t$.
\end{theorem}

In light of the theorem above, it is natural to investigate how the geometry of a fraction determines its optimality. For saturated fractions, some experiments have been presented in \cite{rimini2014d}, where the problem of finding $D$-optimal saturated fractions is translated into an optimization problem using two different objective functions. In words, such objective functions consider the cardinality of the intersection between a fraction and each circuit, and for each circuit they count how many points are needed to complete the circuit. We will recall some results in that direction in the next section.

\section{Design of the simulation study}
\label{sec:2}

In this section we show how fractions generated by the procedure {\tt Proc Optex} can be classified according to their geometrical structure and their $D$-optimality.

To measure the $D$-optimality of a fraction ${\mathcal F}$ we use the $D$-efficiency, see \cite{man3}. The determinant of the information matrix is $D_{\mathcal F} = \det(X_{\mathcal F}^T X_{\mathcal F})$, where $X_{\mathcal F}$ is the model matrix restricted to the fraction points. The $D$-efficiency of ${\mathcal F}$ is then defined as
\[
E_{\mathcal F} = 100 \left( \frac{1}{{\#{\mathcal F}}}
D_{\mathcal F}^{{1}/{\#{\mathcal F}}} \right)
\]
where $\#{\mathcal F}$ is the number of points of ${\mathcal F}$.

To analyze the position of the design points with respect to the supports of the circuits, let us give some definitions. Let $C_A=(c_{ij}, i=1,\ldots,L, j=1,\ldots,K)$ be the matrix, whose rows contain the values of the
indicator functions of the circuits $f_1, \ldots, f_L$, $c_{ij}=(f_{ij} \ne 0) , i=1,\ldots,L, j=1,\ldots,K$ and $Y_{\mathcal F}=((y_{\mathcal F})_1,\ldots,(y_{\mathcal F})_K)$ be the $K$-dimensional column vector that contains the unknown values of the indicator function of the points of ${\mathcal F}$, and let $b=(b_1, \ldots, b_L)$ be the column vector defined by $b_i=\#\mathrm{supp}(f_i), i=1,\ldots , L$.

For each circuit $f_i, i=1,\ldots,L$ we consider the cardinality $(b_{\mathcal F})_i$ of the intersection between its support $\mathrm{supp}(f_i)$ and the fraction ${\mathcal F}$:
\[
(b_{\mathcal F})_i = \langle {\mathrm{supp}}(f_i), Y_{\mathcal F} \rangle \, .
\]
For each fraction ${\mathcal F}$, these value form the vector $b_{\mathcal F} = ((b_{\mathcal F})_1, \ldots, (b_{\mathcal F})_L))$.

In the case of saturated fractions, in \cite{rimini2014d} the geometry of a fraction ${\mathcal F}$ has been summarized through its indicator function $Y$ by means of the objective functions
\[ 
g_2({\mathcal F}) = \sum_{i=1}^L (b-b_{\mathcal F})_i^2 \qquad  \mbox{and}  \qquad 
g_3({\mathcal F}) = \max(b_{\mathcal F}) \, .
\]

In the examples illustrated in \cite{rimini2014d} concerning saturated fractions, the $D$-optimality is reached when the values of $g_2$ and $g_3$ are maximal. This seems to suggest that $D$-optimal fractions correspond to fractions as close as possible to the circuits.

When analyzing fractions with an arbitrary number of points (not necessarily saturated), the objective functions $g_2$ and $g_3$ defined above have a less clear meaning. In fact, for saturated fractions the vector $(b-b_{\mathcal F})$ is strictly positive in view of Theorem \ref{mainthm}, while this property does not hold in general. Another issue which makes the interpretation of $g_2$ and $g_3$ not easy to understand is the fact that there are circuits with different cardinalities.

To overcome the difficulties mentioned above, we have considered here only a subset of the circuits, namely the circuits with support on $4$ points. In the combinatorial theory of contingency tables such simple circuits are known as {\it basic moves} and have several interesting properties, see \cite{haraetal:09} and \cite{rapallo|yoshida:10}. In particular, under mild conditions, the basic moves preserve the connectivity of the fiber of a contingency table without the use of a Markov basis, and nevertheless their number is dramatically smaller than the cardinality of the whole circuit basis. For instance, in the $2^5$ design with main effects there are $353,616$ circuits, but only $720$ of them are basic moves. We denote with ${\overline{\mathcal C}}_A$ the set of the basic moves in ${\mathcal C}_A$, and with $\overline{L}$ its cardinality.

When considering only the basic moves in ${\overline{\mathcal C}}_A$, we can consider only a sub-vector $\overline{b}_{\mathcal F} = ((\overline{b}_{\mathcal F})_1, \ldots, (\overline{b}_{\mathcal F})_{\overline{L}})$ of $b_{\mathcal F}$ using only the basic moves in $\overline{\mathcal C}_A$. Note that by construction the vector $\overline{b}_{\mathcal F}$ has elements in $\{0,1,2,3,4\}$. To analyze a fraction we then consider:
\begin{itemize}
\item the table of counts of $\overline{b}_{\mathcal F}$;

\item the mean and the variance of $\overline{b}_{\mathcal F}$:
\[
\mathrm{m}(\overline{b}_{\mathcal F}) = \frac 1 {\overline{L}} \sum_{i=1}^{\overline{L}} (\overline{b}_{\mathcal F})_i \qquad  \mbox{and} \qquad 
\mathrm{var}(\overline{b}_{\mathcal F}) = \frac 1 {\overline{L}} \sum_{i=1}^{\overline{L}} (\overline{b}_{\mathcal F})^2_i - \mathrm{m}(\overline{b}_{\mathcal F})^2 \, .
\]
\end{itemize}

We have considered the main-effect model for $4$ different designs and with different numbers of design points. More precisely, we have considered the $2^4$ design, the $2^5$ design, the $3^3$ design, and the $2 \times 3 \times 4$ design. For each design, we have considered fractions with $k=p+1,p+2,p+3$ design points, where $p$ is the cardinality of a saturated design or, equivalently, the number of parameters of the model. In all cases we have analyzed $500$ fractions generated by the {\tt Proc Optex}.

\section{Results} \label{sec:3}

\subsection{First scenario. Design $2^4$.}
\label{sec:3.1}

Let us consider first the $2^4$ design. The design matrix $X$ of the full design has $16$ rows and $5$ columns, the number of estimable parameters. The matrix $X$ has rank $5$, and therefore we analyze fractions with $k=6, 7 , 8$ points. For this design, the circuit basis has $1,348$ elements with $100$ basic moves. The remaining $1,248$ circuits have support on $5$ or $6$ points.

We generated $500$ fractions with {\tt Proc Optex} for each of the design sizes $k=6, 7 , 8$, and for each fraction ${\mathcal F}$ we computed the intersections with the $100$ basic moves in $\overline{\mathcal C}_A$, obtaining the vectors $\overline{b}_{\mathcal F}$. We have classified the table of counts of $\overline{b}_{\mathcal F}$, together with its mean and variance, with respect to the $D$-optimality, and the results are reported in Table \ref{tab:des_2_4}.

\begin{table}  \caption{The tables of counts of $\overline{b}_{\mathcal F}$, the means ${\mathrm m}(\overline{b}_{\mathcal F})$, the variances ${\mathrm var}(\overline{b}_{\mathcal F})$, and the $D$-optimality for the $2^4$ design.}
\label{tab:des_2_4}       
%
%
\begin{tabular}{p{1.2cm}p{0.2cm}p{0.55cm}p{0.55cm}p{0.55cm}p{0.55cm}p{0.55cm}p{0.2cm}p{1cm}p{0.2cm}p{1cm}p{0.2cm}p{0.9cm}p{0.2cm}p{0.6cm}}
\hline\noalign{\smallskip}
$\#{\mathcal F}$ & & \multicolumn{5}{c}{table$(\overline{b}_{\mathcal F})$} & & $\mathrm{m}(\overline{b}_{\mathcal F})$ & & $\mathrm{var}(\overline{b}_{\mathcal F})$ & & $E_{\mathcal F}$ & & $n$ \\
\noalign{\smallskip}\hline\noalign{\smallskip}
 & & $0$ & $1$  & $2$ &  $3$ & $4$ & & & & &  & &  \\
\noalign{\smallskip}\hline\noalign{\smallskip}
$k=6$ & & $9$ & $39$ & $45$ & $7$ & $0$ & & $1.5$ & & $0.57$ & & $91.98$ & & $500$  \\
\noalign{\smallskip}\hline\noalign{\smallskip}
$k=7$ & & $6$ & $22$ & $66$ & $3$ & $3$ & & $1.75$ & & $0.55$ & & $93.93$ & & $500$  \\
\noalign{\smallskip}\hline\noalign{\smallskip}
$k=8$ & & $3$ & $18$ & $58$ & $18$ & $3$ & & $2$ & & $0.60$ & & $94.41$ & & $117$  \\
      & & $6$ & $0$ & $88$ & $0$ & $6$ & & $2$ & & $0.48$ & & $100.00$ & & $383$  \\
\noalign{\smallskip}\hline\noalign{\smallskip}
\end{tabular}
\end{table}

The first row of Table \ref{tab:des_2_4} says that all the $500$ fractions with $6$ points generated by {\tt Proc Optex} have a common behavior in terms of intersections with the circuits, namely each fraction has null intersection with $9$ circuits and  intersection on $1$ point with $39$ circuits, on $2$ points with $45$ circuits, on $3$ points with $7$ circuits, while no intersection on $4$ points occur.

In particular for all the fractions sizes we see that, for a given value of $E_{\mathcal F}$, all the fractions have exactly the same vector $\overline{b}_{\mathcal F}$. In all the remaining scenarios we will observe that, for a given value of $E_{\mathcal F}$ all the fractions have just few possible values of $\overline{b}_{\mathcal F}$.  This confirms once more the connection between $D$-optimality and circuits.

For $k=8$ points {\tt Proc Optex} provides $117$ fractions with $E_{\mathcal F}=94.41$ and $387$ fractions with $E_{\mathcal F}=100$ (these latter ones are resolution III orthogonal designs). Both groups of designs have the same mean value of $\overline{b}_{\mathcal F}$ (equal to $2$) but different variances ($0.60$ and $0.48$). The designs with the highest value of $E_{\mathcal F}$ have the lowest variance, that is $0.48$.

\subsection{Second scenario. Design $2^5$.}
\label{sec:3.2}

We analyze now the $2^5$ design. The design matrix $X$ of the full design has $32$ rows and $6$ columns, the number of estimable parameters. The matrix $X$ has rank $6$, and therefore we analyze fractions with $k=7, 8 , 9$ points. For this design, the circuit basis has $353,616$ elements, but there are only $720$ basic moves.

We generated $500$ fractions with {\tt Proc Optex} for each of the sample sizes $k=7, 8 , 9$, and the results of the simulation study are reported in Table \ref{tab:des_2_5}.

\begin{table}  \caption{The tables of counts of $\overline{b}_{\mathcal F}$, the means ${\mathrm m}(\overline{b}_{\mathcal F})$,  the variances ${\mathrm var}(\overline{b}_{\mathcal F})$, and the $D$-optimality for the $2^5$ design.}
\label{tab:des_2_5}       
%
%
\begin{tabular}{p{1.2cm}p{0.2cm}p{0.55cm}p{0.55cm}p{0.55cm}p{0.55cm}p{0.55cm}p{0.2cm}p{1cm}p{0.2cm}p{1cm}p{0.2cm}p{0.9cm}p{0.2cm}p{0.6cm}}
\hline\noalign{\smallskip}
$\#{\mathcal F}$ & & \multicolumn{5}{c}{table$(\overline{b}_{\mathcal F})$} & & $\mathrm{m}(\overline{b}_{\mathcal F})$ & & $\mathrm{var}(\overline{b}_{\mathcal F})$ & & $E_{\mathcal F}$ & & $n$ \\
\noalign{\smallskip}\hline\noalign{\smallskip}
 & & $0$ & $1$  & $2$ &  $3$ & $4$ & & & & &  & &  \\
\noalign{\smallskip}\hline\noalign{\smallskip}
$k=7$ & & $249$ & $321$ & $141$ & $9$ & $0$ & & $0.88$ & &  $0.58$ & & $83.87$ & & $12$  \\
      & & $238$ & $342$ & $132$ & $8$ & $0$ & & $0.88$ & & $0.54$ & & $88.18$ & & $20$  \\
      & & $230$ & $353$ & $135$ & $1$ & $1$ & & $0.88$ & & $0.51$ & & $90.71$ & & $468$  \\
\noalign{\smallskip}\hline\noalign{\smallskip}
$k=8$ & & $194$ & $348$ & $162$ & $16$ & $0$ & & $1$ & & $0.58$ & & $90.86$ & & $35$  \\
      & & $191$ & $344$ & $182$ & $0$ & $3$ & & $1$ & & $0.56$ & & $95.32$ & & $182$  \\
      & & $186$ & $352$ & $180$ & $0$ & $2$ & & $1$ & & $0.53$ & & $100.00$ & & $283$  \\
\noalign{\smallskip}\hline\noalign{\smallskip}
$k=9$ & & $157$ & $335$ & $212$ & $13$ & $3$ & & $1.12$ & & $0.61$ & & $95.10$ & & $82$  \\
      & & $155$ & $339$ & $209$ & $15$ & $2$ & & $1.12$ & & $0.60$ & & $97.58$ & & $418$  \\
\noalign{\smallskip}\hline\noalign{\smallskip}
\end{tabular}
\end{table}

For all the fraction sizes that we have considered in this scenario, {\tt Proc Optex} provides groups of designs with different values of the $D$-optimality criterion. We observe that for each fraction size the mean of $\overline{b}_{\mathcal F}$ is constant while the variances of $\overline{b}_{\mathcal F}$ decrease as $E_{\mathcal F}$ increase. This result could suggest that the algorithm for searching for $D$-optimal designs could be improved if also the variances were taken into account.

\subsection{Third scenario. Design $3^3$.}
\label{sec:3.3}

We consider a multilevel design, namely the $3^3$ design. The design matrix $X$ of the full design has $27$ rows and $7$ columns, the number of estimable parameters. The matrix $X$ has rank $7$, and therefore we analyze fractions with $k=8, 9 , 10$ points. For this design, the circuit basis has $73,071$ elements, $243$ of which are basic moves.

We generated $500$ fractions with {\tt Proc Optex} for each of the sample sizes $k=8, 9 , 10$, and the results of the simulation study are reported in Table \ref{tab:des_3_3}.

\begin{table}  \caption{The tables of counts of $\overline{b}_{\mathcal F}$, the means ${\mathrm m}(\overline{b}_{\mathcal F})$,  the variances ${\mathrm var}(\overline{b}_{\mathcal F})$, and the $D$-optimality for the $3^3$ design.}
\label{tab:des_3_3}       
%
%
\begin{tabular}{p{1.2cm}p{0.2cm}p{0.55cm}p{0.55cm}p{0.55cm}p{0.55cm}p{0.55cm}p{0.2cm}p{1cm}p{0.2cm}p{1cm}p{0.2cm}p{0.9cm}p{0.2cm}p{0.6cm}}
\hline\noalign{\smallskip}
$\#{\mathcal F}$ & & \multicolumn{5}{c}{table$(\overline{b}_{\mathcal F})$} & & ${\mathrm m}(\overline{b}_{\mathcal F})$ & &  $\mathrm{var}(\overline{b}_{\mathcal F})$ & & $E_{\mathcal F}$ & & $n$ \\
\noalign{\smallskip}\hline\noalign{\smallskip}
 & & $0$ & $1$  & $2$ &  $3$ & $4$ & & & & &  & &  \\
\noalign{\smallskip}\hline\noalign{\smallskip}
$k=8$ & & $48$ & $108$ & $81$ & $6$ & $0$ & & $1.19$ & & $0.60$ & & $52.59$ & & $19$  \\
      & & $40$ & $122$ & $77$ & $4$ & $0$ & & $1.19$ & & $0.51$ & & $55.72$ & & $185$  \\
      & & $39$ & $120$ & $84$ & $0$ & $0$ & & $1.19$ & & $0.47$ & & $56.67$ & & $296$  \\
\noalign{\smallskip}\hline\noalign{\smallskip}
$k=9$ & & $29$ & $112$ & $94$ & $8$ & $0$ & & $1.33$ & & $0.53$ & & $57.95$ & & $192$  \\
      & & $27$ & $108$ & $108$ & $0$ & $0$ & & $1.33$ & & $0.44$ & & $62.45$ & & $308$  \\
\noalign{\smallskip}\hline\noalign{\smallskip}
$k=10$ & & $21$ & $96$ & $114$ & $12$ & $0$ & & $1.48$ & & $0.52$ & & $61.02$ & & $500$  \\
\noalign{\smallskip}\hline\noalign{\smallskip}
\end{tabular}
\end{table}

As in the previous scenarios for each fraction size all the designs have the same mean of $\overline{b}_{\mathcal F}$ while the best designs have the lowest variances of $\overline{b}_{\mathcal F}$.

\subsection{Fourth scenario. Design $2 \times 3 \times 4$.}
\label{sec:3.4}

The last scenario concerns an asymmetric design, the $2 \times 3 \times 4$ design.  The design matrix $X$ of the full design has $24$ rows and $7$ columns, the number of estimable parameters. The matrix $X$ has rank $7$, and therefore we analyze fractions with $k=8, 9 , 10$ points. For this design, the circuit basis has $13,470$ elements, $174$ of which are basic moves.

\begin{table}  \caption{The tables of counts of $\overline{b}_{\mathcal F}$, the means ${\mathrm m}(\overline{b}_{\mathcal F})$, the variances ${\mathrm var}(\overline{b}_{\mathcal F})$, and the $D$-optimality for the $2 \times 3 \times 4$ design.}
\label{tab:des_2_3_4}       
%
%
\begin{tabular}{p{1.2cm}p{0.2cm}p{0.55cm}p{0.55cm}p{0.55cm}p{0.55cm}p{0.55cm}p{0.2cm}p{1cm}p{0.2cm}p{1cm}p{0.2cm}p{0.9cm}p{0.2cm}p{0.6cm}}
\hline\noalign{\smallskip}
$\#{\mathcal F}$ & & \multicolumn{5}{c}{table$(\overline{b}_{\mathcal F})$} & & $\mathrm{m}(\overline{b}_{\mathcal F})$ & & $\mathrm{var}(\overline{b}_{\mathcal F})$ & & $E_{\mathcal F}$ & & $n$ \\
\noalign{\smallskip}\hline\noalign{\smallskip}
 & & $0$ & $1$  & $2$ &  $3$ & $4$ & & &  & &  & &  \\
\noalign{\smallskip}\hline\noalign{\smallskip}
$k=8$ & & $20$ & $80$ & $70$ & $4$ & $0$ & & $1.33$ & & $0.50$ & & $51.71$ & & $332$  \\
      & & $21$ & $76$ & $76$ & $0$ & $1$ & & $1.33$ & & $0.50$ & & $51.71$ & & $168$  \\
\noalign{\smallskip}\hline\noalign{\smallskip}
$k=9$ & & $14$ & $68$ & $84$ & $7$ & $1$ & & $1.50$ & & $0.53$ & & $52.80$ & & $246$  \\
      & & $14$ & $69$ & $81$ & $10$ & $0$ & & $1.50$ & & $0.53$ & & $52.80$ & & $254$  \\
\noalign{\smallskip}\hline\noalign{\smallskip}
$k=10$ & & $8$ & $60$ & $88$ & $18$ & $0$ & & $1.67$ & & $0.52$ & & $54.25$ & & $157$  \\
       & & $9$ & $56$ & $94$ & $14$ & $1$ & & $1.67$ & & $0.52$ & & $54.25$ & & $286$  \\
       & & $11$ & $48$ & $106$ & $6$ & $3$ & & $1.67$ & & $0.52$ & & $54.25$ & & $57$  \\
\noalign{\smallskip}\hline\noalign{\smallskip}
\end{tabular}
\end{table}

The results of this scenario, displayed in Table \ref{tab:des_2_3_4}, reinforce the connection between $D$-optimality and the variance of $\overline{b}_{\mathcal F}$. For each fraction size we obtain two groups of designs with different $\overline{b}_{\mathcal F}$ but equal means ${\mathrm m}(\overline{b}_{\mathcal F})$, equal variances ${\mathrm var}(\overline{b}_{\mathcal F})$ and equal $D$-efficiencies $E_{\mathcal F}$.

\section{Concluding remarks}
\label{sec:conclusion}

The simulations discussed in the previous sections for various designs show that the
cardinalities of the intersections between a fraction and the basic moves are able to predict the $D$ optimality of the fraction. In particular, as low is the variance of such cardinalities as high is the $D$-efficiency of the fraction, at least for the simple-effect models analyzed here.

These results are encouraging and suggest to analyze such connection in a more general framework, in order to characterize $D$-optimal fractions following three main directions: first, to study the behavior of the $D$-optimality in terms of the intersection with the basic moves also for models with interactions, and to find connections with other known notions, such as uniformity and discrepancy; second, to characterize the basic moves in order to extend our study to large-sized designs; finally, to implement the criterion based on the basic moves in statistical software to improve the existing algorithm for finding $D$-optimal designs.

\bibliographystyle{spmpsci}
\bibliography{referenc_ROBERTOFONTANA}

\end{document}